\documentclass[%
 reprint,
 amsmath, amssymb,
 aps,
]{revtex4-2}
\usepackage{graphicx}
\usepackage{dcolumn}
\usepackage{bm}
\usepackage{hyperref}
\usepackage{amsmath}
\usepackage{nameref}
\usepackage{enumitem}
\usepackage{url}
\usepackage{gensymb}
\usepackage{setspace}

\usepackage{xcolor}

\begin{document}

\preprint{APS/123-QED}

\title{Newton’s First Law Is Not a Special Case of the Second Law}

\author{Indresh Yadav$^{1, 2}$}
\altaffiliation{indresh@iitbbs.ac.in} \altaffiliation{indresh@mit.edu}
\author{P. M. Geethu$^{3}$}
\altaffiliation{geethukrishnanpm@gmail.com}

\affiliation{$^1$ Department of Physics, Indian Institute of Technology Bhubaneswar, Odisha  752050, India\\
$^2$  Department of Chemical Engineering, Massachusetts Institute of Technology, Cambridge, Massachusetts 02139, United States\\
$^3$ Department of Physics, National Institute of Technology Calicut, Kerala  673601, India}%

\begin{abstract}
Newton's Laws of Motion form the basis of classical mechanics, but misconceptions about their interrelationships persist in pedagogy. A prevalent misunderstanding is that Newton’s First Law is a trivial consequence of the Second Law. This paper argues that the First Law serves a logically distinct foundational role that defines the context in which the Second Law is valid. This conceptual distinction is clarified through classical thought experiments and further supported by insights from relativistic mechanics. Furthermore, the paper discusses the notion of the zeroth Law. It evaluates several candidates, including the absoluteness of space and time, the conservation and additivity of mass, and the locality of force in time. By articulating the details of the logical structure of Newton's Laws, this article offers theoretical clarity and pedagogical value for the teaching and interpretation of Newtonian mechanics.

\end{abstract}

\maketitle

\section{Introduction} 
In Aristotelian physics, motion was categorized as either natural or violent  \cite{rovelli2015aristotle, drake1964galileo}. Natural motion referred to the spontaneous movement of bodies toward their elemental destinations, for example, stones falling toward the Earth, whereas violent motion results from external forces such as pushing or throwing. Based on observations of bodies moving through resistive media like air and water, the Aristotelian view assumed that continuous force is needed to maintain motion. 
Furthermore, the study of dynamics in this framework was grounded in qualitative reasoning rather than in a precise mathematical formulation. This perspective began to shift with the pioneering work of Galileo, who introduced the principle of inertia \cite{drake1964galileo}. His insights laid the conceptual foundation for Newton’s First Law, which redefined motion as a condition that persists in the absence of a net external force and requires force only to change the state of motion of an object. Subsequently, Newton \cite{newton1934principia}  and others \cite{pourciau2020principia, coelho2018deduction}, including Leibniz, Euler, and Laplace, formalized these ideas into a rigorous mathematical framework now known as Newton’s Laws of Motion. This framework enables us to describe and predict the trajectories and accelerations of bodies under the influence of forces, and thereby the time evolution of dynamical systems.

Despite the apparent simplicity of Newton’s Laws, significant misconceptions persist in their pedagogical interpretation \cite{rigden1987high, anderson1990newton, pfister2004newton}. Among the most prevalent is the belief that Newton’s First Law is merely a special case of the Second Law when the net force is zero \cite{rigden1987high}. This article aims to address and clarify such misunderstandings by examining the distinct and complementary conceptual roles that each law plays within the framework of Newtonian mechanics.  

In addition to revisiting the logical structure of Newton’s three formal laws, this paper explores the deeper, often implicit assumptions on which Newtonian mechanics is based. The first among these is the nature of mass \cite{wilczek2006origin}, the foundational assumption that mass is a conserved, additive, and invariant quantity. Although often taken for granted, this premise underlies the applicability of Newton’s Second Law. From both a logical and a pedagogical point of view, this assumption merits recognition as a candidate for Newton's Zeroth Law. We also discuss two additional candidates for the Zeroth Law, the classical assumption of absolute space and time \cite{sep-newton-stm}, and the principle of instantaneous response to force \cite{scherr2005newton}, that a body responds only to the net force acting on it at that instant. These principles serve as metaphysical or operational prerequisites for the laws of motion to function meaningfully.

Although Newtonian mechanics provides a highly successful description of macroscopic phenomena at low speeds, it has been superseded in certain regimes by modern physical theories. Einstein’s theory of relativity \cite{grunbaum1955logical} reinterpreted space and time as a single interconnected continuum whose measurements depend on the observer’s state of motion, thus replacing Newton's concept of absolute simultaneity\textsuperscript{\ref{note: a}}. Similarly, quantum mechanics reveals that motion at the microscopic scale is not governed by deterministic trajectories but by probabilistic wavefunctions \cite{ballentine1970statistical}. Nonetheless, Newtonian mechanics remains a limiting case of these broader frameworks and continues to serve as a conceptual and pedagogical foundation. A clearer understanding of its logical structure not only strengthens its instructional value but also deepens appreciation of how Newtonian mechanics connects with and diverges from modern physics.

\section{Newton’s Laws of Motion}

\subsection{Newton’s First Law} 
Newton's First Law of Motion, commonly known as the Principle of Inertia, traces its conceptual origin to the work of Galileo Galilei \cite{drake1964galileo}. Galileo used thought experiments to explore the ideas of inertia and relative motion. One of his well-known examples involves a smoothly sailing ship \cite{galilei2023dialogue}.
He imagined observers confined to a cabin below deck, performing various experiments such as dropping objects or observing flying insects, and argued that they would be unable to determine whether the ship was at rest or moving at a constant velocity. Galileo reasoned that everything inside the cabin, including the air, shares the motion of the ship. As a result, the outcomes of the experiments would appear identical whether the ship was stationary or moving uniformly (i.e., at a constant speed in a straight line). This led him to conclude that uniform motion is indistinguishable from rest without reference to the external world, an insight that challenged the Aristotelian idea of absolute rest. This insight laid the foundation for what Newton later formalized as the First Law of Motion stated as \cite{ newton1934principia}

First Law- \textit{Every body continues in its state of rest, or of uniform motion in a
right line, unless it is compelled to change that state by forces impressed upon it.}

This is often paraphrased in modern textbooks as

\textit{An object at rest remains at rest, and an object in motion continues in uniform motion unless acted upon by an external force.}

This law states that an object will maintain a constant velocity if the net external force acting on it is zero. In other words, either no force is applied, or all forces acting on the object cancel out exactly. Any deviation from constant velocity, such as a change in speed or direction, requires a non-zero net force, indicating an interaction with the environment.

More precisely, Newton’s First Law articulates the physical indistinguishability between rest and uniform motion in a straight line. This is not a matter of perception but a reflection about the symmetry of physical laws. The outcomes of all mechanical\textsuperscript{\ref{note: b}} experiments remain unchanged under a transformation from rest to uniform motion, revealing that such states are physically indistinguishable. This invariance\textsuperscript{\ref{note: c}} forms the basis for defining an inertial frame of reference coordinate system in which an object moves with constant velocity in the absence of force. Newton’s First Law, therefore, does not simply describe passive motion; it provides the operational standard for identifying inertial frames. In classical mechanics, all inertial frames are physically equivalent, and none is fundamentally preferred over the others. 

\subsection{Newton’s Second Law} 
In the framework of Newtonian physics, force\textsuperscript{\ref{note: d}} is the agent responsible for changes in an object’s motion, and these changes are quantified through the concept of momentum. The relationship between force and the rate of change of momentum defines how motion evolves under external influences. This is formalized in Newton’s Second Law of Motion, which states \cite{newton1934principia}.

Second Law- \textit{A change in motion is proportional to the motive force impressed and takes place along the straight line in which that force is impressed.}

Suppose a force \( \vec{F} \) is applied to a body over a time interval \( \Delta t \). The effect of this force over time is described by a vector quantity known as impulse, denoted by \( \vec{J} \). This impulse produces a change in the momentum \( \vec{P} \) of the body.
\begin{equation}
\vec{J} = \vec{F} \, \Delta t = \Delta \vec{p}
\end{equation}
Building on this, Newton, his contemporary and successor, including Leibniz, Euler, d'Alembert, Lagrange, and Laplace, extended the concept to forces applied continuously over time rather than impulsively \cite{pourciau2020principia,pourciau2011newton, nauenberg2012comment, coelho2018deduction}. This leads to the differential form of the equation of motion, expressing Newton’s Second Law as a limit process 
\begin{equation}
\vec{F} = \lim_{\Delta t \to 0} \frac{\Delta \vec{p}}{\Delta t} = \frac{d\vec{p}}{dt}
\end{equation} 
In Newtonian mechanics, the mass $m$ of an object is considered constant in time regardless of velocity \( \vec{v} \). Under this assumption, the equation simplifies to its most familiar form 
\begin{equation}
\vec{F} = m \frac{d\vec{v}}{dt} = m \vec{a} 
\label{eq:eq3}
\end{equation}
 Equation \eqref{eq:eq3} connects three conceptually distinct quantities, force \(\vec{F}\), representing the influence exerted by the external environment, mass $m$, an intrinsic property of the object, and acceleration $a$, which characterizes the movement of the object through space and time. The real strength of the differential form of Newton's Second Law lies in its ability to use instantaneous\textsuperscript{\ref{note: e}} information, such as instantaneous position and velocity, to predict the continuous long-term behavior of a dynamical system.

\subsection{Newton’s Third Law}  
In Newtonian mechanics, the forces exerted by two objects on each other are inherently linked, following a precise symmetry. This symmetry is fundamental to the conservation of linear momentum. The nature of these mutual forces is described by Newton's Third Law, stated as \cite{newton1934principia}

Third Law- \textit{To every action, there is always an equal and opposite reaction, or the mutual interactions between two bodies are always equal in magnitude and directed in opposite directions.}

\begin{figure}[h]
\centering
\includegraphics[width=0.7\linewidth]{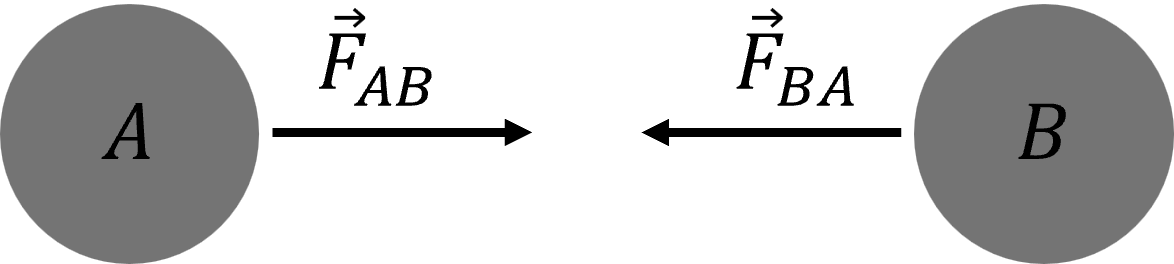}
\caption{Illustration of Newton's Third Law: Body \( B \) exerts a force \( \vec{F}_{AB} \) on body \( A \), and body \( A \) simultaneously exerts an equal and opposite force \( \vec{F}_{BA} = -\vec{F}_{AB} \) on body \( B \). These forces form an \textit{action-reaction} pair, acting on different bodies but always equal in magnitude and opposite in direction.}
    \label{fig:fig1}
\end{figure}
Mathematically, it can be expressed as
\begin{equation}
\vec{F}_{AB} = -\vec{F}_{BA}
\label{eq:eq4}
\end{equation}
where \( \vec{F}_{AB} \) is the force exerted on body \( A \) by body \( B \), and \( \vec{F}_{BA} \) is the force exerted on body \( B \) by body \( A \) (Figure-\ref{fig:fig1}). These forces are equal in magnitude but opposite in direction, and they act on different bodies. 

In its classical formulation, Newton’s Third Law requires that the forces two bodies exert on each other are not only equal in magnitude and opposite in direction, but also \textit{collinear}, that is, they act along the same line connecting the centers of the interacting bodies. This requirement holds in many cases, such as gravitational and electrostatic interactions between point particles, or idealized contact forces under rigid-body assumptions. This is characteristic of central forces. However, this directional symmetry does not hold universally. In systems involving non-central forces, such as magnetic interactions between moving charges, the action-reaction pair may still be equal in magnitude but can fail to be collinear or exactly opposite in direction, leading to a breakdown of Newton’s Third Law in its strict classical form.

\begin{figure}[h]
\centering
\includegraphics[width=0.97\linewidth]{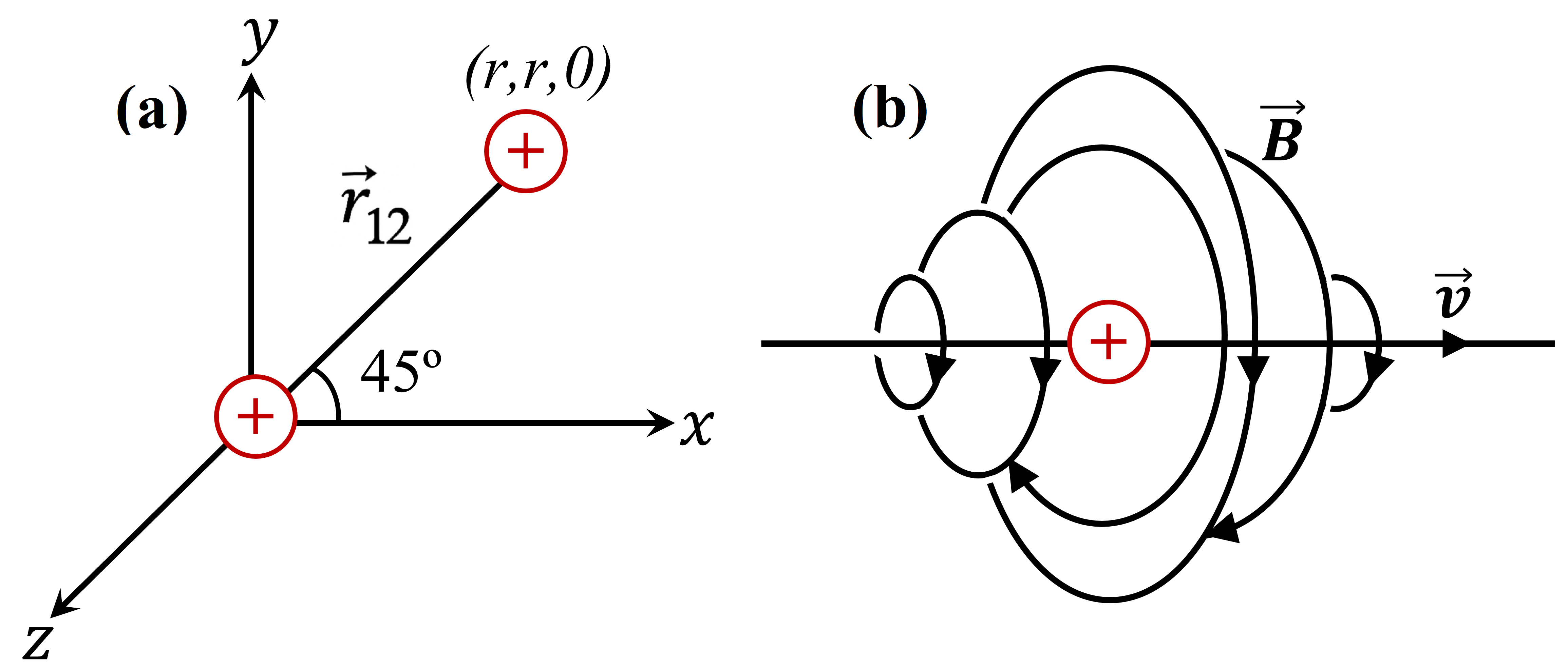}
\caption{(a) Two protons in a right-handed coordinate system. One located at the origin \((0, 0, 0)\), and the other at \((r, r, 0)\). The separation vector \(\vec{r}_{12}\) lies in the \(xy\)-plane at a 45\degree{} angle to the \(x\)-axis, indicating their relative positions. (b) Magnetic field lines produced by a proton moving at constant velocity along the \(x\)-axis.
}
    \label{fig:fig2}
\end{figure}

This phenomenon can be exemplified by two protons moving with identical velocities along the \( +x \)-axis within a right-handed coordinate system (Figure-\ref{fig:fig2}(a)). Let proton 1 be located at the origin \( (0, 0, 0) \), and proton 2 at position \( (r, r, 0) \) at a given instant. The separation vector lies in the \( xy\) plane at a 45\degree angle, given by \( \vec{r}_{12} = r\hat{x} + r\hat{y} \). Each moving proton generates a magnetic field that circulates around its velocity vector according to the right-hand rule \cite{jefimenko1994direct, griffiths2023introduction}. At the location of proton 2, the magnetic field due to proton 1 points in the \( +z \)-direction (Figure-\ref{fig:fig2}(b)). By the Lorentz force law, \( \vec{F} = e(\vec{v} \times \vec{B}) \), where $e$ is the electric charge of a proton, with \( \vec{v} \parallel \hat{x} \) and \( \vec{B} \parallel \hat{z} \), the resulting force on the proton 2 points along \( -\hat{y} \). In contrast, the magnetic field at proton 1 due to proton 2 points in the \( -z \)-direction, resulting in a force on proton 1 along \( +\hat{y} \). Although these forces are equal in magnitude and opposite in direction, they are not collinear, as they do not act along the separation vector, which lies diagonally in the \( xy \)-plane.  

In a different scenario, consider proton 1 moving along the \( +x \)-axis and proton 2 moving along the \( -y \)-axis. In this configuration, the magnetic field at proton 2 due to proton 1 points along the \( +\hat{z} \)-direction, and similarly, the field at proton 1 due to proton 2 also points along \( +\hat{z} \). The magnetic force on proton 1 acts along \( -\hat{y} \), while the force on proton 2 acts along \( -\hat{x} \). In addition to the magnetic force between protons, there is Coulombic repulsion that acts along the line connecting them and satisfies Newton’s Third Law. 
However, the magnetic forces are not aligned along this line and therefore do not obey the geometric form of Newton's Third Law. These magnetic forces are of the order \( v^2/c^2 \) relative to the Coulomb force, where $c$ is the speed of light. Such effects become significant only when the particles move at speeds close to the speed of light. 

The greatest significance of Newton’s Third Law lies not in the action-reaction symmetry itself, but in its profound implications, most notably the law of conservation\textsuperscript{\ref{note: f}} of linear momentum. This conservation holds under the condition that the system is closed or isolated, meaning no external forces are acting on it \cite{feynman1965feynman}.

Consider two bodies, \( A \) and \( B \), as shown in Figure~\ref{fig:fig1}. In a closed system, the force exerted on each body by the other is equal to the time rate of change of its own momentum, given as
\begin{equation}
\vec{F}_{AB} = \frac{d\vec{P}_A}{dt}, \quad \vec{F}_{BA} = \frac{d\vec{P}_B}{dt}
\label{eq:eq5}
\end{equation}
Now, by Newton’s Third Law, these forces are equal in magnitude and opposite in direction; thus, using Equations \eqref{eq:eq4} and \eqref{eq:eq5}, we can write 
\begin{equation}
\frac{d\vec{P}_A}{dt} = -\frac{d\vec{P}_B}{dt}
\quad \Rightarrow \quad \frac{d}{dt}(\vec{P}_A + \vec{P}_B) = 0
\end{equation}
This means that the total momentum of the system is constant over time, i.e.,
\begin{equation}
\vec{P}_A + \vec{P}_B = \text{constant}
\end{equation} 
Note that since Equation \eqref{eq:eq5} is nothing but Newton's Second Law, confusion also arises that Newton's Third Law is a special case of the Second Law \cite{gangopadhyaya2017can}. However, the Second Law does not inform anything about the nature of the mutual forces exerted by $A$ and $B$ on each other. In contrast, the Third Law specifies the symmetry of these mutual interactions. Thus, the conservation of linear momentum is the consequence of the combined features of the Second and Third Laws. 

This conservation principle holds even in modern physics, including quantum theory and relativity, where Newton's Third Law in its classical form may no longer be valid. Despite this, momentum is still conserved in such systems, but conservation must be understood in a broader context. For example, in electrodynamics, the electromagnetic field carries momentum, and thus the total momentum of the system comprising both particles and fields is conserved.

\section{Newton’s First Law: Not a Special Case of the Second Law} A common misconception is that Newton's First Law is merely a special case of Newton’s Second Law. The usual argument proceeds as follows.
\begin{equation}
\vec{F} = m \frac{d\vec{v}}{dt}
\end{equation}
If the net force acting on a body is zero, i.e., \( \vec{F} = 0 \), then
\begin{equation}
\frac{d\vec{v}}{dt} = 0
\end{equation}
Integrating this with respect to time yields
\begin{equation}
\vec{v}(t) = \vec{C}
\end{equation}
Here, \( \vec{C} \) is an integration constant which can be determined from the initial conditions. It may represent zero or any constant velocity, depending on the initial state of motion of the body. Hence, the velocity remains constant in the absence of external forces. This interpretation makes it seem as though the First Law is redundant. However, this is not the case for the following reasons.

\textbf{Defining Inertial Frames}: 
In mechanics, a reference frame provides the geometric structure necessary to describe motion by defining positions, velocities, and accelerations \cite{sep-spacetime-iframes}. However, to attribute clear dynamical meaning to motion, that is, to relate it to forces, a special class of frames is required, known as inertial frames \cite{clemence1966inertial}. An inertial frame is one in which a body remains at rest or moves with constant velocity unless acted upon by a force. This condition is formalized in Newton’s First Law, which thereby serves to define inertial frames. The Second Law of motion, which relates force to acceleration, is valid only within such frames. Without the First Law to provide a criterion for identifying inertial frames, the Second Law would lose its physical significance, as it would be impossible to distinguish whether an observed acceleration arises from a genuine force or from the acceleration of the reference frame itself. 

\begin{figure}[h]
\centering
\includegraphics[width=0.85\linewidth]{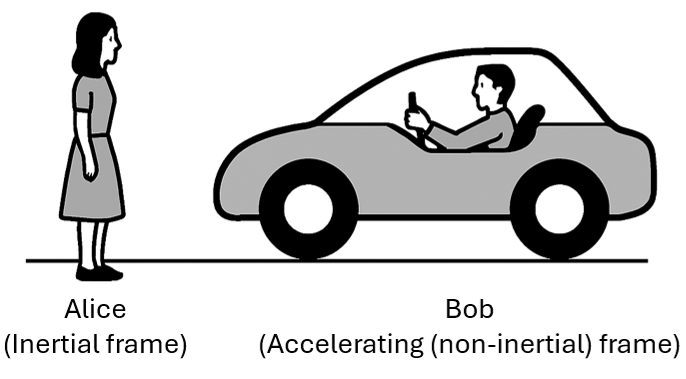}
\caption{Alice and Bob: Empirical observations are consistent with Newton’s Second Law in an inertial frame (Alice’s frame), but appear to violate it in a non-inertial (accelerating) frame (Bob’s frame). This illustrates the necessity of identifying inertial frames for the correct application of Newton’s Second Law.}
    \label{fig:fig3}
\end{figure}
To illustrate this, consider two observers: Alice and Bob. Alice is standing on the ground, which we approximate as an inertial frame, while Bob is inside a car that begins to accelerate (Figure-\ref{fig:fig3}). From Alice’s perspective, Bob is accelerating due to the force exerted by the car’s engine. She can correctly apply Newton’s Second Law to describe Bob’s acceleration by accounting for the net external forces acting on him, including the propulsive force of the car’s engine and the opposing force of friction. However, from Bob's perspective inside the accelerating car, it appears that Alice and the external world accelerate backwards, despite no observable force acting on them.  If Bob were to apply Newton's Second Law to account for Alice's acceleration, it would yield incorrect results. Alice appears to accelerate despite the fact that no observable force is acting on her. This apparent failure of the Second Law arises because Bob is in a non-inertial (accelerating) frame of reference.

This example highlights the operational significance of Newton’s First Law. It identifies inertial frames, those in which the Second Law holds in its standard form. Without this criterion, there would be no consistent method for determining whether a particular frame of reference is suitable for applying Newton’s Second Law. 

\textbf{Support from Relativistic Mechanics}:
Importantly, Newton’s First Law remains unchanged even in the framework of special relativity. In special relativity, inertial frames are still defined as those in which the First Law holds true. Although the transformation between inertial frames is no longer Galilean but Lorentzian, the fundamental principle that uniform motion is physically indistinguishable from rest remains intact.  
In contrast, the classical form of the Second Law, \( \vec{F} = m\vec{a} \), does not hold exactly in special relativity, since mass depends on velocity and force is not necessarily parallel to acceleration. Instead, the relativistically consistent form \( \vec{F} = \frac{d\vec{p}}{dt} \), with momentum defined as \( \vec{p} = \gamma m \vec{v} \), where \( \gamma = \frac{1}{\sqrt{1 - v^2/c^2}} \), remains valid. While this relativistic formulation remains valid, it presupposes the identification of inertial frames and thus relies on Newton's First Law to define them. The fact that, in the absence of force, relativistic momentum and hence velocity remain constant should not be interpreted as a derivation of the First Law from the Second Law. Rather, the First Law defines the conditions under which all dynamical laws, including the relativistic form of the Second Law, hold.

Thus, Newton’s First Law retains its fundamental and logically independent status even beyond Newtonian mechanics. It is not a mere special case of the Second Law, but a foundational empirical principle that delineates the domain in which the Second Law is valid. In this sense, it fulfills both an operational and an empirical-axiomatic role within the logical structure of classical physics. 

\section{Newton’s Zeroth Law}
While Newton did not explicitly formulate a Zeroth Law, his mechanics rest on several deep assumptions that precede and support the three formal laws of motion. These foundational ideas, though often implicit, are crucial to the logical structure and practical application of Newtonian physics. 

One such assumption is the concept of absolute space and time, the idea that space and time exist as immutable backgrounds, independent of any observer. It can be summarized as follows\\
\textit{Space and time provide an absolute stage in which bodies move and interact, independent of the observer’s motion.}\\
In Newton's view, space and time are distinct and ontologically real entities \cite{sep-newton-stm}. Space exists independently of the material bodies it contains and provides a universal, immovable frame in which all physical processes occur. Time flows uniformly and independently of any observer's state of motion, proceeding at a constant rate throughout the universe. This axiom underpinned Newton's worldview and the formulation of his laws \cite{anderson1990newton}. However, empirical evidence and philosophical developments \cite{sep-spacetime-holearg} suggest that motion is only meaningful in a relative sense, that is, we can only observe changes in an object’s motion with respect to other objects. Despite this, Newton's mathematical formalism remains consistent and effective even when interpreted within a relative framework of space and time. Therefore, while conceptually elegant, absolute space and time are not strictly necessary for the validity of Newtonian mechanics. 

A second foundational assumption concerns the locality and instantaneity of force stated as \cite{scherr2005newton}\\
\textit{At any instant of time, an object responds only to the forces it experiences at that instant.}\\
This principle is embedded in the differential form of Newton's Second Law, Equation \eqref{eq:eq3}. It states that a body’s response is determined by the instantaneous net force acting upon it, and that no memory of past forces is required. This locality in time is essential for setting up well-defined initial value problems in Newtonian mechanics.

A third foundational assumption in Newtonian mechanics is the nature of mass \cite{wilczek2006origin}, which can be summarized as\\
\textit{Mass is an additive, invariant\textsuperscript{\ref{note: c}} and conserved quantity. In a closed system, the total mass equals the sum of the masses of its constituent parts and remains constant over time.}\\
This principle is deeply embedded in Newtonian mechanics. It enables the use of Newton's Second Law, ensures consistency in momentum conservation, and underlies the dynamics of interacting systems. It is so fundamental that it often goes unmentioned in textbooks. Yet without it, the logical structure of Newtonian mechanics would be incomplete. 

Among the above three foundational assumptions, the additivity and invariance of mass stand out as the most essential and universally valid within Newtonian theory. It supports the entire formal structure of Newtonian dynamics. Although, special relativity later challenged the conservation of mass as an absolute truth, within the Newtonian framework, it remains a natural and essential assumption and arguably deserving the title of Newton’s Zeroth Law. Framing it as a Zeroth Law not only clarifies the implicit assumptions underlying Newton’s formal laws but also enhances the pedagogical presentation of Newtonian mechanics by making its logical structure more explicit.

\vspace{-0.7em}

\section{Summary}
This paper reexamines the foundational structure of Newtonian mechanics, questioning the common pedagogical assumption that Newton’s First Law is merely a limiting case of the Second Law. We argue that the First Law serves a distinct and indispensable role by defining inertial reference frames, the context within which the Second Law holds. Furthermore, we discuss and suggest a candidate for Newton’s Zeroth Law. 

Taken together, the Zeroth, First, Second, and Third Laws form a logically coherent and hierarchically ordered structure. The Zeroth Law establishes the foundational assumption that mass is an additive, invariant, and conserved quantity. This assumption underpins the applicability of Newton’s dynamical laws and provides the ontological foundation for studying the mechanics of physical systems. The First Law defines the class of reference frames, namely inertial frames. It is important to note that Newton’s First Law is not a special case of the Second Law, but a foundational empirical principle that delineates the domain in which the Second Law is valid. The Second Law then provides the quantitative relationship between force and the rate of change of momentum. Finally, the Third Law introduces the principle of mutual interactions and leads directly to the conservation of linear momentum. This paper underscores the importance of conceptual clarity in both the teaching and theoretical interpretation of Newtonian mechanics. A careful distinction between the assumptions and the operational laws not only aids in pedagogy, but also reinforces the internal logical structure of Newtonian physics.

\section*{Notes and Glossary}
\begin{enumerate}[label=\textsuperscript{\alph*}, left=0pt, itemsep=0pt]
\item \label{note: a} In Newtonian mechanics, simultaneity is absolute; if two events occur at the same time in one inertial frame, they are assumed to occur at the same time in all inertial frames. In contrast, the theory of relativity treats simultaneity as relative; two events that are simultaneous in one frame may not be simultaneous in another, depending on the observer's state of motion.

\item \label{note: b} In the framework of special relativity, this extends to all physical experiments, where the laws of physics are invariant under transformations between inertial frames of reference. 
\item \label{note: c} Invariance refers to a quantity remaining unchanged under a specified transformation, such as a change in coordinate system or reference frame. In Newtonian mechanics, mass is invariant under Galilean transformations, meaning that all inertial observers measure the same value.
\item \label{note: d} An impressed force is an external influence applied to a body to change its state of motion, that is, to initiate motion from rest or to alter uniform motion in a straight line.
\item \label{note: e} Instantaneous velocity is rigorously defined in calculus as the limit of the average velocity as the time interval approaches zero. However, physical measurements of velocity necessarily involve at least two distinct position-time data points taken over a finite interval. Therefore, instantaneous velocity is fundamentally a mathematical abstraction that represents the rate of change in position at a single moment in time. In practical applications, what we refer to as instantaneous velocity is always an approximation, valid only when the measurement interval is sufficiently small that variations in velocity during that interval can be neglected.
\item \label{note: f} In physics, a conservation law asserts that a particular measurable quantity remains unchanged over time in an isolated system, however, internal processes or interactions may occur. Although energy, momentum, or other properties can be redistributed among parts of the system, the total amount of each conserved quantity remains constant. 


\end{enumerate}

\vspace{1em}

\bibliography{apssamp}

\end{document}